\documentclass[%
 reprint,
 superscriptaddress,
preprintnumbers,
 amsmath,amssymb,
prd,
longbibliography
]{revtex4-2}

\usepackage{graphicx}% Include figure files
\usepackage{bm}% bold math
\usepackage{hyperref}% add hypertext capabilities
\usepackage[utf8]{inputenc}
\usepackage{siunitx}
\usepackage{orcidlink}

\newcommand{\rmd}{\mathrm{d}}
\newcommand{\rmi}{\mathrm{i}}
\newcommand{\vv}{\vec{v}}
\newcommand{\vu}{\vec{u}}

\newcommand{\hrho}{\hat{\rho}}
\newcommand{\hchi}{\hat{\chi}}
\newcommand{\heta}{\hat{\eta}}
\newcommand{\hbeta}{\hat{\beta}}
\newcommand{\hsigma}{\hat{\sigma}}
\newcommand{\hH}{\hat{H}}  
\newcommand{\hV}{\hat{V}}

\newcommand{\idty}{\hat{\mathbb{I}}}

\newcommand{\GF}{G_\mathrm{F}}
\newcommand{\ket}[1]{\left| #1 \right\rangle}
\newcommand{\avg}[1]{\left\langle #1 \right\rangle}
\DeclareMathOperator{\tr}{tr}

\newcommand{\figscale}{1}

\begin{document}

\preprint{INT-PUB-24-006}
\preprint{LA-UR-24-21029}

\title{Once-in-a-lifetime encounter models for neutrino media:\\
From coherent oscillations to flavor equilibration}

\newcommand{\LANL}{Theoretical Division, Los Alamos National Laboratory, Los Alamos, New Mexico, 87545}
\newcommand{\UNM}{Department of Physics and Astronomy, University of New Mexico, Albuquerque, New Mexico 87131, USA}

\author{Anson Kost\,\orcidlink{0009-0008-0327-5057}}

% \author{Nishant Raina}
\affiliation{\UNM}

\author{Lucas Johns}
\affiliation{\LANL}

\author{Huaiyu Duan\,\orcidlink{0000-0001-6708-3048}}
\affiliation{\UNM}

\date{\today}

\begin{abstract}
    Collective neutrino oscillations are typically studied using the lowest-order quantum kinetic equation, also known as the mean-field approximation. However, some recent quantum many-body simulations suggest that quantum entanglement among neutrinos may be important and may result in flavor equilibration of the neutrino gas. In this work, we develop new quantum models for neutrino gases in which any pair of neutrinos can interact at most once in their lifetimes. A key parameter of our models is $\gamma=\mu \Delta z$, where $\mu$ is the neutrino coupling strength, which is proportional to the neutrino density, and $\Delta z$ is the duration over which a pair of neutrinos can interact each time. Our models reduce to the mean-field approach in the limit $\gamma\to0$ and achieve flavor equilibration in time $t \gg (\gamma\mu)^{-1}$. These models demonstrate the emergence of coherent flavor oscillations from the particle perspective and may help elucidate the role of quantum entanglement in collective neutrino oscillations.
\end{abstract}
\maketitle

\section{Introduction}

Flavor oscillations in a dense neutrino gas can be described by the quantum kinetic equation which, at the lowest order, gives the so-called ``mean-field approximation''
\begin{align}
    (\partial_t + \vv\cdot\vec{\nabla}) \varrho = -\rmi [\mathsf{H}, \varrho].
    \label{eq:kinetic}
\end{align}
Here $\varrho(t,\vec{r}, \vec{p})$ is the Wigner distribution of the neutrino field of momentum $\vec{p}$ and velocity $\vv=\vec{p}/p$ at time $t$ and position $\vec{r}$, and $\mathsf{H}$ is the Hamiltonian that dictates the coherent flavor evolution of the neutrino field \cite{Sigl:1993ctk}. Although there have been some debates about the adequacy of this one-particle picture in the past (e.g., \cite{Bell:2003mg,Friedland:2003eh}), most of the literature on collective neutrino oscillations has adopted the mean-field approach (see, e.g., \cite{Duan:2010bg,Chakraborty:2016yeg,Tamborra:2020cul,Volpe:2023met} for reviews). However, some recent many-body calculations suggest that the quantum entanglement among neutrinos can be important for neutrino oscillations and may even result in flavor equilibration (but not necessarily flavor equipartition) (e.g., \cite{Cervia:2019res,Patwardhan:2021rej,Illa:2022zgu,Martin:2023ljq,Martin:2023gbo}; see \cite{Patwardhan:2022mxg} for a review; see also \cite{Shalgar:2023ooi,Johns:2023ewj} for some counterarguments). 

One of the issues with existing many-body neutrino models is that they involve a closed system of a limited number of neutrinos that interact with each other indefinitely. In a realistic astrophysical environment, such as a core-collapse supernova (CCSN) or a binary-neutron star merger (BNSM), a few neutrinos may interact for a brief moment while their wave packets overlap, and those neutrinos may never see each other again. With this in mind, we develop new quantum models in which any pair of neutrinos may have a once-in-a-lifetime encounter (OILE). In the OILE models, each neutrino is treated as an open quantum system with the rest of the neutrino gas as its environment.

The limited goal of this paper is not to compute neutrino oscillations in a realistic CCSN or BNSM environment nor to resolve all the questions about the mean-field approach and the existing many-body models, but rather to demonstrate how the results of these models can appear as appropriate limits in the same quantum model.

% We adopt the natural units $\hbar=c=1$ throughout the paper.

\section{A foreground neutrino through a uniform medium}
\label{sec:medium}

We first consider a model in which a neutrino passes through a uniform and dense neutrino medium. We assume that the foreground neutrino travels along the $z$ axis and encounters a distinct background neutrino in each spatial interval $[z_{n-1}, z_n)$ where $z_n=n\Delta z$ with $\Delta z$ being a constant and $n$ a natural number. Inside the $n$th interval, the foreground and background neutrinos co-evolve with a constant Hamiltonian $\hH_n$ so that
\begin{align}
    \hrho (t) = e^{-\rmi\hH_n (t-t_{n-1})} \hrho_{n-1} e^{\rmi\hH_n (t-t_{n-1})},
    % \qquad(t_{n-1}\leq t < t_n),
    \label{eq:evolution}
\end{align} 
where $t_{n-1}=z_{n-1}$, and $\hrho$ is the two-neutrino density operator.

A key assumption of the OILE models is that the two neutrinos participating in the interaction will not be further correlated through future interactions and will not be measured against each other for correlation. With this assumption, the correlation between the foreground and background neutrinos in the $(n-1)$th interval has left the foreground neutrino in a mixed state at the end of the interval which is described by the one-body density operator
\begin{align}
    \hchi_{n-1} &= \tr_\eta [\hrho(t_{n-1})],
\end{align}
where $\tr_\eta$ denotes the partial trace over the background neutrino. At the beginning of the $n$th interval, the two-body density operator of the foreground neutrino and another background neutrino is simply
\begin{align}
    \hrho_{n-1} &= \hchi_{n-1} \otimes \heta,
\end{align}
where $\heta$ is the one-body density operator of the background neutrino before the interaction. 

We employ the two-flavor mixing scheme (between $\nu_e$ and $\nu_x$) for simplicity. In this scheme, the one-body density operators can be expressed in terms of the identity operator $\idty$, the Pauli operators $\hsigma_{j}$ ($j=1,2,3$), and the Bloch vectors $\bm{P}$ and $\bm{Q}$ as
\begin{align}
    \hchi = \frac{1}{2} (\idty + \bm{P}\cdot\bm{\hsigma}) 
    \quad\text{and}\quad
    \heta = \frac{1}{2} (\idty + \bm{Q}\cdot\bm{\hsigma}).
\end{align}
The purity of the foreground neutrino is measured by $|\bm{P}|$, which is also a measure of how strongly it is correlated with the background neutrinos.

For the two-neutrino interaction, we adopt the constant (forward-scattering) interaction Hamiltonian 
\begin{subequations}
    \label{eq:V}
    \begin{align}
        \hV 
        &= \frac{\mu}{2}(1-\vv\cdot\vu) \, \hsigma_j\otimes\hsigma_j
        \label{eq:V-a}
        \\
        %  (\bm{\hsigma}\odot\bm{\hsigma})\\
        &=  \frac{\mu}{2}(1-\vv\cdot\vu)
        \begin{pmatrix}
            1 & & & \\
            & -1 & 2 & \\
            & 2 & -1 & \\
            & & & 1
        \end{pmatrix}
    \end{align}        
\end{subequations}
between the foreground and background neutrinos, where $\mu$ is the interaction strength, and $\vv$ and $\vu$ are the velocities of the foreground and background neutrino respectively. Einstein's summation over repeated indices is assumed in Eq.~\eqref{eq:V-a} and throughout this paper.

The interaction Hamiltonian in Eq.~\eqref{eq:V} is the same as that in \cite{Friedland:2003eh} (up to a trace term) if the interaction strength is defined as $\mu=\sqrt{2}\GF/\mathcal{V}$, where $\GF$ is the Fermi coupling constant and $\mathcal{V}$ is the normalization volume. Although it is not entirely clear what $\mathcal{V}$ should be in a general scenario, it seems natural to take $\mathcal{V}=\Delta z^3$ in this model with $\Delta z$ representing the size of the neutrino wave packet. This implies $\mu^{-1} \sim \SI{0.1}{cm}$ for $\Delta z \sim \SI{e-11}{cm}$ \cite{Kersten:2015kio}, and 
\begin{align}
    \Vert\hrho_{n+1}-\hrho_{n}\Vert \propto \gamma \equiv \mu\Delta z \sim \SI{e-10}{}.
    \label{eq:drho}
\end{align}

\begin{figure}[htb]
    \includegraphics[trim=1 1 1 1, clip, scale=\figscale]{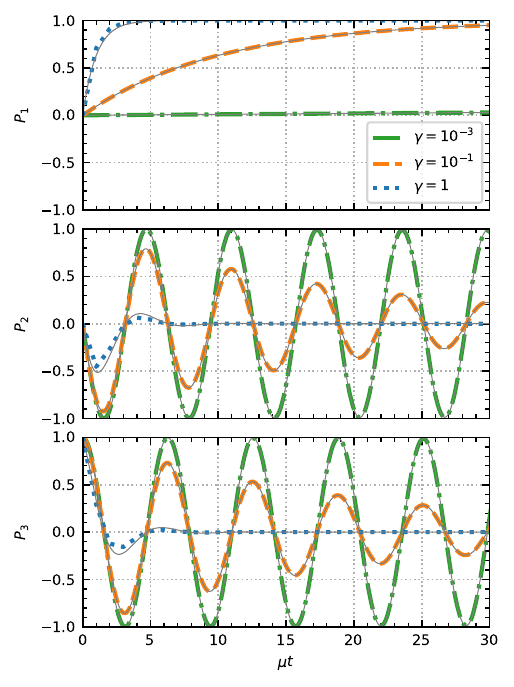}
    \caption{The three components of the (flavor-basis) Bloch vector of the foreground neutrino at $t=n\Delta z$ for the three different interaction interval sizes $\Delta z=\gamma/\mu$ (as labeled). The foreground neutrino is initially in $\ket{\nu_e}$, and the background neutrinos are in $(\ket{\nu_e} + \ket{\nu_x})/\sqrt2$ before the interaction. The thin solid curves represent the solutions to Eq.~\eqref{eq:master-P}.}
    \label{fig:P-bg}
\end{figure}

To highlight the effect of the neutrino medium, we ignore the vacuum Hamiltonian by setting $\hH_n = \hV$ for now. We also assume $\vv\cdot\vu=0$ for simplicity. As a concrete example, consider the case where the foreground neutrino has pure electron flavor at $t=z=0$, and the background neutrinos are all in the quantum flavor state $(\ket{\nu_e} + \ket{\nu_x})/\sqrt2$ before interacting with the foreground neutrino. To demonstrate the effect of small $\gamma$, we employ three artificially large $\gamma$ values ($\gamma=1$, $10^{-1}$, and $10^{-3}$) and compute $\bm{P}(t_n)$ with these values. The results are shown as dotted, dashed, and dot-dashed curves in Fig.~\ref{fig:P-bg}. For the cases where $\gamma \ll 1$, Fig.~\ref{fig:P-bg} shows the coherent oscillation of the foreground neutrino induced by the neutrino medium with a period of $2\pi/\mu$. This figure also shows that the flavor of the foreground neutrino approaches that of the medium on a time scale $\gtrsim (\mu\gamma)^{-1}$. 

The above results can be understood through the following master equation which approximately describes the flavor evolution of the foreground neutrino  (see Appendix~\ref{sec:master}):
\begin{subequations}
    \label{eq:master-P}
    \begin{align}
        \dot{\bm{P}} 
            &\approx\mu\bm{Q}\times\bm{P} - \gamma\mu(\bm{P} - \bm{Q})
            -\frac{\gamma\mu}{2}\bm{Q}\times(\bm{Q}\times\bm{P})
            \label{eq:master-P-a}\\
            &=\mu\bm{Q}\times\bm{P} - \gamma\mu\left[\bm{P}_\parallel
            + \left(1-\frac{|\bm{Q}|^2}{2}\right)\bm{P}_\perp\right]
            + \gamma\mu\bm{Q},
            \label{eq:master-P-b}
    \end{align}
\end{subequations}
where $\bm{P}_\parallel$ and $\bm{P}_\perp$ are the components of $\bm{P}$ that are parallel and perpendicular to $\bm{Q}$, respectively. We plot the solutions to Eq.~\eqref{eq:master-P} as thin solid curves in Fig.~\ref{fig:P-bg}. They agree with the simulation results for $\gamma \ll 1$ and $\mu t\leq 30$.

The first term on the right-hand side of Eq.~\eqref{eq:master-P-b} describes the coherent flavor oscillation of the foreground neutrino induced by the neutrino medium. It is the same as the neutrino-neutrino interaction term in the mean-field equation \eqref{eq:kinetic} if one defines $\mu=\sqrt2\GF n_\nu$ with $n_\nu$ being the number density of background neutrinos. This is indeed true from the perspective of the foreground neutrino in our model because it encounters one background neutrino in each volume $\mathcal{V}$. The second term in Eq.~\eqref{eq:master-P-b} describes the decoherence of the foreground neutrino due to the interaction with the background neutrinos. The third term has the opposite effect and increases the coherence of the foreground neutrino when the neutrino medium has a net flavor polarization. The net effect of the second and third terms is that the flavor of the foreground neutrino approaches that of the medium on the timescale $(\gamma\mu)^{-1}$.

\section{Collective oscillations and decoherence}

Having understood the effect of a uniform neutrino medium in the previous model, we now would like to see how collective oscillations can emerge from a similar model. Because it is unrealistic to keep track of all neutrinos in a CCSN or BNSM environment, we consider an ensemble of $N$ ``particles'' evolving in discrete time steps with step size $\Delta z$. Each particle in the ensemble represents a group of neutrinos with similar initial conditions. We again assume that each pair of \emph{physical} neutrinos interact with each other at most once in their lifetimes, and we define the total density operator of the ensemble at the beginning of the $n$th time step as
\begin{align}
    \hrho_{n-1} = \bigotimes_{a=1}^N \hchi_{n-1}(a),
\end{align}
where $\hchi_{n-1}(a)$ is the one-body density operator of the $a$th particle at the end of the previous step. The evolution of the ensemble during the $n$th step is still given by Eq.~\eqref{eq:evolution}, but now with
\begin{align}
    \hH_n = -\sum_{a=1}^N \frac{\omega_a}{2}\hbeta_a + \sum_{a<b} \Theta_{ab}(n) \hV_{ab}.
    \label{eq:H}
\end{align}
Here $\omega_a$ is the vacuum oscillation frequency of the $a$th particle, and
\begin{align}
    \hbeta_a = \left(\bigotimes_{b\neq a}\idty(b)\right)\otimes\hsigma_3(a)
\end{align}
in the mass basis, where $\hsigma_j(a)$ and $\idty(a)$ are the Pauli and identity operators for the $a$th particle, respectively. Also in Eq.~\eqref{eq:H}, $\Theta_{ab}(n)=1$ if the particles $a$ and $b$ interact in the $n$th time step and 0 otherwise, and 
\begin{align}
    \hV_{ab} = \frac{\mu}{2}(1-\vv_a\cdot\vv_b)
    \left(\bigotimes_{c\neq a,b}\idty(c)\right) \otimes
    \hsigma_j(a)\otimes\hsigma_j(b).
\end{align}

\begin{figure*}[thb]
    \includegraphics[trim=1 1 1 1, clip, scale=\figscale]{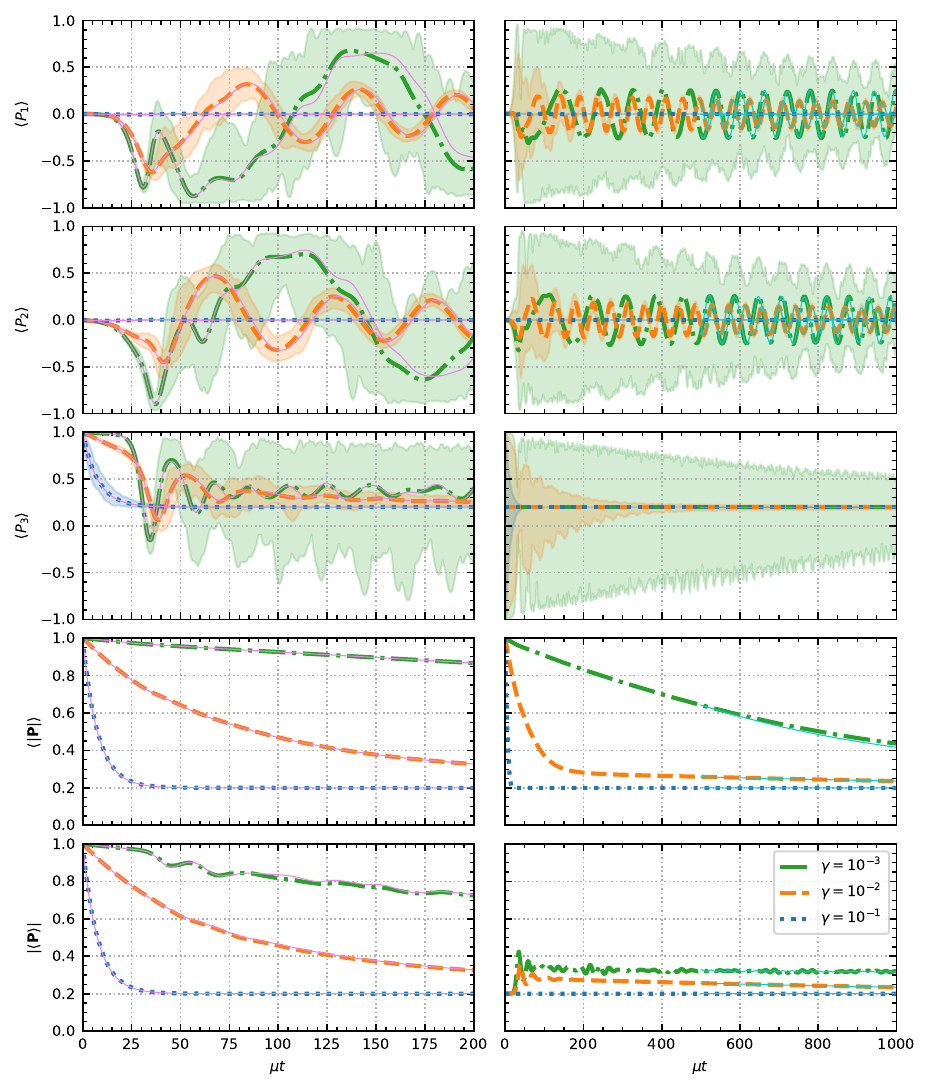}
    \caption{The evolution of an ensemble of 100 particles with 60 $\nu_e$'s and 40  $\nu_x$'s initially. The dotted, dashed, and dot-dashed curves represent the mean quantities in the mass basis (labeled by vertical axes) for the cases with $\gamma=10^{-1}$, $10^{-2}$, and $10^{-3}$, respectively, and the shadows in the top three panels represent the ranges of the corresponding quantities. The quantities in the left panels are averaged over the particles that start as $\nu_e$, and the right panels show the averages over the entire ensemble. The thin solid curves in the left and right panels represent the solutions to Eqs.~\eqref{eq:master-P-int} and \eqref{eq:master-P-int-avg}, respectively.}
    \label{fig:P-int}
\end{figure*}

As an example, we consider an ensemble of $N=100$ particles with random but fixed velocity directions. The ensemble has a ``bipolar'' initial condition: one group of 60 particles are initially in $\ket{\nu_e}$ and have the same vacuum oscillation frequency $\omega=0.1\mu$, and the other 40 particles start as $\nu_x$ with $\omega=0.2\mu$. A small effective mixing angle $\theta=10^{-3}$ is used to mimic the effect of the matter background \cite{Hannestad:2006nj}. In each time step, all particles undergo one interaction in random, mutually exclusive pairs. The evolution of the ensemble is calculated with 3 different step sizes that correspond to $\gamma=10^{-1}$, $10^{-2}$, and $10^{-3}$, respectively. In Fig.~\ref{fig:P-int} (left column, top three panels) we show the average Bloch vectors $\avg{\bm{P}}$ for the first group of particles as dotted, dashed, and dot-dashed curves and their ranges as shadows. We also show the magnitudes of the average Bloch vectors $|\avg{\bm{P}}|$ as well as the averaged magnitudes of the Bloch vectors $\avg{|\bm{P}|}$ for the first particle group in the same figure (left column, two bottom panels).

The system with $\gamma=10^{-3}$ demonstrates a flavor-pendulum-like evolution \cite{Hannestad:2006nj} for $\mu t\lesssim 30$. Afterwards, the Bloch vectors $\bm{P}_a$ of the individual particles with different velocities $\vv_a$ diverge from each other. This results in kinematic decoherence, which is manifested as decreasing $|\avg{\bm{P}}|$ \cite{Raffelt:2007yz}. In contrast, the evolution of the system with $\gamma=10^{-1}$ is dominated by quantum decoherence, which shrinks the magnitudes of individual Bloch vectors $|\bm{P}_a|$ before the flavor pendulum falls. In general, kinematic decoherence is bounded by quantum decoherence because
\begin{align}
    |\avg{\bm{P}}| \leq \avg{|\bm{P}|}.
\end{align}
The results of the simulation with $\gamma=10^{-2}$ lie between these two cases.

The numerical results can be understood through the following master equation:
\begin{widetext}
\begin{align}
    \dot{\bm{P}}_a
    &\approx (-\omega_a\bm{B} + \mu\avg{(1-\vv_a\cdot\vv_b)\bm{P}_b}_b)\times\bm{P}_a
    -\gamma\mu\left[
        \avg{(1-\vv_a\cdot\vv_b)^2}_b \bm{P}_a
        -\avg{(1-\vv_a\cdot\vv_b)^2 \bm{P}_b}_b
    \right]
    \nonumber\\
    &\quad
    -\frac{\gamma\mu}{2}[
        \avg{(1-\vv_a\cdot\vv_b)\bm{P}_b}_b\times(\avg{(1-\vv_a\cdot\vv_c)\bm{P}_c}_c\times\bm{P}_a)
        + \avg{(1-\vv_a\cdot\vv_b)\avg{(1-\vv_b\cdot\vv_c)(\bm{P}_c\times\bm{P}_b)}_c}_b\times\bm{P}_a
    ],
    \label{eq:master-P-int}
\end{align}
% \end{widetext}
where $\bm{B}=(0,0,1)$ in the mass basis, and 
\begin{align}
    \avg{(1-\vv_a\cdot\vv_b)^s \bm{P}_b}_b &= \frac{1}{N-1}\sum_{b\neq a} (1-\vv_a\cdot\vv_b)^s \bm{P}_b 
    % \\
    % &
    \xrightarrow{N\gg1} \frac{1}{N} \sum_{b} (1-\vv_a\cdot\vv_b)^s \bm{P}_b.
\end{align}
\end{widetext}
We show the solutions to Eq.~\eqref{eq:master-P-int} as the thin solid curves in the left panels of Fig.~\ref{fig:P-int}. They agree with the simulation results reasonably well considering the random nature of the model.

Because the neutrino-neutrino interaction Hamiltonian is non-integrable \cite{Martin:2023gbo}, the evolution of the individual Bloch vectors is somewhat randomized and is not correlated with the neutrino velocities after a sufficiently long time. Using the ansatz
\begin{align}
    \sum_{b} (1-\vv_a\cdot\vv_b)^s \bm{P}_b
    &\approx
    \avg{\bm{P}}\sum_{b} (1-\vv_a\cdot\vv_b)^s ,
\end{align}
we rewrite Eq.~\eqref{eq:master-P-int} as
\begin{align}
    \dot{\bm{P}}_a
    &\approx(-\omega_a\bm{B} + \mu\avg{\bm{P}})\times\bm{P}_a
    -\frac{4}{3}\gamma\mu(\bm{P}_a - \avg{\bm{P}})
    \nonumber\\
    &\quad
    -\frac{\gamma\mu}{2}\avg{\bm{P}}\times(\avg{\bm{P}}\times\bm{P}_a)
    \label{eq:master-P-int-avg}
\end{align}
when the Bloch vectors are not correlated with the velocity vectors, where $\avg{\cdots}$ represents the average over the whole ensemble. The above equation looks very much like Eq.~\eqref{eq:master-P-a} except for the vacuum oscillation term, $\bm{Q}\to\avg{\bm{P}}$, and a factor of
\begin{align}
    \frac{1}{N}\sum_b (1-\vv_a\cdot\vv_b)^2 \approx \frac{4}{3}
\end{align}
in the second term on the right-hand side of the equation. 

In the right panels of Fig.~\ref{fig:P-int} we show the results for the same calculations as those in the left panels but over a longer duration and for the quantities averaged over the whole ensemble instead of the first group only. We also use the Bloch vectors in the simulations at $t\mu=500$ as initial conditions and solve Eq.~\eqref{eq:master-P-int-avg}. The results are shown as thin solid curves in the same panels and are in reasonable agreement with the simulations.

Equation~\eqref{eq:master-P-int-avg} shows that significant quantum decoherence can occur at $t\gtrsim (\gamma\mu)^{-1}$, which is confirmed by Fig.~\ref{fig:P-int}. Unlike the previous model where the foreground neutrino approaches the flavor state of the constant background medium, Fig.~\ref{fig:P-int} shows that quantum decoherence drives the system toward the fixed point:
\begin{align}
    \bm{P}_a \xrightarrow{\gamma \mu t\gg1} (0, 0, \avg{P_3}),
\end{align}
where $\avg{P_3}$ is a constant of motion of both Eqs.~\eqref{eq:master-P-int} and \eqref{eq:master-P-int-avg}. However, for $t\lesssim (\gamma\mu)^{-1}$, the system is well described by the mean-field approximation, which is equivalent to setting $\gamma=0$ in Eq.~\eqref{eq:master-P-int}.

Naively, one may think that kinematic decoherence will completely destroy collective oscillations at the mean-field level. However, Fig.~\ref{fig:P-int} (right panels) shows that $\avg{\bm{P}}$ can precess around $\bm{B}$ for a sustained period of time when $\gamma\ll1$. This collective precession of the Bloch vectors is known as flavor synchronization \cite{Pastor:2001iu} and can be understood as follows. 

Summing Eq.~\eqref{eq:master-P-int-avg} for all particles, we obtain
\begin{align}
    \dot{\bm{J}}
    &\approx -\bm{B}\times\bm{M},
    \label{eq:J-M}
\end{align}
where
\begin{align}
    \bm{J} = \sum_{a=1}^N \bm{P}_a
    \quad\text{and}\quad
    \bm{M} = \sum_{a=1}^N \omega_a\bm{P}_a.
\end{align}
Assume $\gamma\ll\omega/\mu\ll1$. According to Eq.~\eqref{eq:master-P-int-avg}, each Bloch vector precesses around $\avg{\bm{P}}$ or $\bm{J}$ on the short time scale $\mu^{-1}$.  Averaging over this fast motion, one expects $\bm{M}$ to be parallel to $\bm{J}$ so that Eq.~\eqref{eq:J-M} becomes
\begin{align}
    \dot{\bm{J}}
    &\approx -\Omega\bm{B}\times\bm{J}
\end{align}
on the intermediate timescale $\omega^{-1}$, where the synchronization frequency is given by
\begin{align}
    \Omega \approx \frac{\bm{J}\cdot\bm{M}}{|\bm{J}|^2}.
\end{align}
Interestingly, the quantum decoherence of individual neutrinos slows down during the synchronization regime where $\bm{P}_a\approx \avg{\bm{P}}$. [See Eq.~\eqref{eq:master-P-int-avg}; see also the dashed curve in the fourth panel in the right column of Fig.~\ref{fig:P-int}.]

\section{Discussion and conclusion}

We have developed two OILE models to study flavor oscillations in dense neutrino gases. Unlike existing many-body models in literature, the OILE models treat each neutrino as an open quantum system and the rest of the medium as its environment. The results of our models depend critically on the model parameter $\gamma=\mu \Delta z$, where $\mu$ characterizes the strength of the neutrino-neutrino interaction and is proportional to the neutrino density, and $\Delta z$ is the typical interaction duration of two neutrinos. Our models reduce to the mean-field approximation in the limit $\gamma\to0$ and exhibit quantum decoherence at $t\gtrsim (\gamma\mu)^{-1}$.

In the first model, we considered a single neutrino passing through a constant background neutrino medium. This model seems to be well approximated by the master equation \eqref{eq:master-P-b}, which has three terms, each with its own physical interpretation: a coherent term that survives at the mean-field level, a term that reduces the coherence of the foreground neutrino, and a term that drives the foreground neutrino to the quantum flavor state of the medium at $t\gtrsim (\gamma\mu)^{-1}$.

In the second model, we considered an ensemble of particles that initially represent $\nu_e$ and $\nu_x$. For the cases where $\gamma\ll1$, we found that the system behaves like a flavor pendulum before kinematic decoherence takes over. On a longer timescale, the system exhibits flavor synchronization before quantum decoherence eventually drives the system to flavor equilibration.

There are some obvious limitations and potential concerns for our models, some of which we plan to address in future work. For example, we adopt the forward scattering Hamiltonian of the neutrino. In reality, neutrinos can have many different final momentum states after scattering. Additionally, all neutrinos are paired to interact in each time step of fixed duration in our models, whereas in reality, the interaction times are probabilistic in nature. Nevertheless, our models shed light on how coherent neutrino oscillations and quantum decoherence can manifest themselves as different limits of the same quantum model from the particle perspective. These models may also serve as a starting point for future studies of these intriguing phenomena.

\begin{acknowledgments}
    We thank J.~Carlson, I.~Deutsch, A.~Friedland, G.~M.~Fuller, J.~D.~Martin, D.~Neill, N.~Raina, and A.~Roggero for helpful discussions. We also thank the Institute for Nuclear Theory at the University of Washington for its kind hospitality and stimulating research environment where this work was initiated. This research was supported in part by INT's US DOE grant No.\ DE-FG02-00ER41132. A.~K.\ and H.~D.\ are supported by the US DOE NP grant No.\ DE-SC0017803 at UNM. L.~J.\ is supported by a Feynman Fellowship through LANL LDRD project No.\ 20230788PRD1.
\end{acknowledgments}

\appendix
\section{Master equations}
\label{sec:master}
Here we briefly outline the derivation of the master equations used in the paper. Using the identities
\begin{align}
    \hsigma_i \hsigma_j &= \delta_{ij} + \rmi \epsilon_{ijk} \hsigma_k 
    \intertext{and}
    [\hsigma_k\otimes\hsigma_k, \hsigma_i\otimes\hsigma_j] 
    &=2\rmi\epsilon_{ijk}(\idty\otimes\hsigma_k - \hsigma_k\otimes\idty),
\end{align}
one can show that
\begin{widetext}
\begin{subequations}
    \begin{align}
        \hchi_{n+1}
        &= \tr_\eta\left\{
            e^{-\rmi \hV \Delta z}\left[  \hchi_n\otimes\heta \right] e^{\rmi \hV \Delta z} 
        \right\} \\
        &\approx \hchi_n 
        -(\rmi \Delta z)\tr_\eta \{
            [\hV, \hchi_n\otimes\heta]
        \} 
        % \nonumber\\
        % &\quad 
        + \frac{(-\rmi\Delta z)^2}{2}\tr_\eta \{
            [\hV, [\hV, \hchi_n\otimes\heta]]
        \}\\
        &= \hchi_n - (\rmi \mu\Delta z)[\heta, \hchi_n] - (\mu\Delta z)^2(\hchi_n - \heta)
        \label{eq:chi-discrete}
    \end{align}    
\end{subequations}
\end{widetext}
for the one-body density operator of the foreground neutrino in the first OILE model (Sec.~\ref{sec:medium}), where $\hV$ is defined in Eq.~\eqref{eq:V} with $\vv\cdot\vu=0$. Numerically, Eq.~\eqref{eq:chi-discrete} is consistent with the master equation
\begin{align}
    \frac{\rmd\hchi}{\rmd t} 
    &\approx -\rmi \mu[\heta, \hchi] - \gamma\mu (\hchi - \heta)
    + \frac{1}{2}\gamma\mu [\heta, [\heta, \hchi]] 
\end{align}    
up to $\mathcal{O}(\Delta z^2)$, which gives Eq.~\eqref{eq:master-P} for the Bloch vector $\bm{P}$ of the foreground neutrino. Eq.~\eqref{eq:master-P-int} can be derived in a similar way by replacing $\heta$ with the average density operators of the background neutrinos and taking into account the geometry factors $1-\vv_a\cdot\vv_b$ in $\hV$.
% \end{widetext}

\bibliography{ref}% Produces the bibliography via BibTeX.

\begin{thebibliography}{19}%
\makeatletter
\providecommand \@ifxundefined [1]{%
 \@ifx{#1\undefined}
}%
\providecommand \@ifnum [1]{%
 \ifnum #1\expandafter \@firstoftwo
 \else \expandafter \@secondoftwo
 \fi
}%
\providecommand \@ifx [1]{%
 \ifx #1\expandafter \@firstoftwo
 \else \expandafter \@secondoftwo
 \fi
}%
\providecommand \natexlab [1]{#1}%
\providecommand \enquote  [1]{``#1''}%
\providecommand \bibnamefont  [1]{#1}%
\providecommand \bibfnamefont [1]{#1}%
\providecommand \citenamefont [1]{#1}%
\providecommand \href@noop [0]{\@secondoftwo}%
\providecommand \href [0]{\begingroup \@sanitize@url \@href}%
\providecommand \@href[1]{\@@startlink{#1}\@@href}%
\providecommand \@@href[1]{\endgroup#1\@@endlink}%
\providecommand \@sanitize@url [0]{\catcode `\\12\catcode `\$12\catcode `\&12\catcode `\#12\catcode `\^12\catcode `\_12\catcode `\%12\relax}%
\providecommand \@@startlink[1]{}%
\providecommand \@@endlink[0]{}%
\providecommand \url  [0]{\begingroup\@sanitize@url \@url }%
\providecommand \@url [1]{\endgroup\@href {#1}{\urlprefix }}%
\providecommand \urlprefix  [0]{URL }%
\providecommand \Eprint [0]{\href }%
\providecommand \doibase [0]{https://doi.org/}%
\providecommand \selectlanguage [0]{\@gobble}%
\providecommand \bibinfo  [0]{\@secondoftwo}%
\providecommand \bibfield  [0]{\@secondoftwo}%
\providecommand \translation [1]{[#1]}%
\providecommand \BibitemOpen [0]{}%
\providecommand \bibitemStop [0]{}%
\providecommand \bibitemNoStop [0]{.\EOS\space}%
\providecommand \EOS [0]{\spacefactor3000\relax}%
\providecommand \BibitemShut  [1]{\csname bibitem#1\endcsname}%
\let\auto@bib@innerbib\@empty
%</preamble>
\bibitem [{\citenamefont {Sigl}\ and\ \citenamefont {Raffelt}(1993)}]{Sigl:1993ctk}%
  \BibitemOpen
  \bibfield  {author} {\bibinfo {author} {\bibfnamefont {G.}~\bibnamefont {Sigl}}\ and\ \bibinfo {author} {\bibfnamefont {G.}~\bibnamefont {Raffelt}},\ }\bibfield  {title} {\bibinfo {title} {{General kinetic description of relativistic mixed neutrinos}},\ }\href {https://doi.org/10.1016/0550-3213(93)90175-O} {\bibfield  {journal} {\bibinfo  {journal} {Nucl. Phys. B}\ }\textbf {\bibinfo {volume} {406}},\ \bibinfo {pages} {423} (\bibinfo {year} {1993})}\BibitemShut {NoStop}%
\bibitem [{\citenamefont {Bell}\ \emph {et~al.}(2003)\citenamefont {Bell}, \citenamefont {Rawlinson},\ and\ \citenamefont {Sawyer}}]{Bell:2003mg}%
  \BibitemOpen
  \bibfield  {author} {\bibinfo {author} {\bibfnamefont {N.~F.}\ \bibnamefont {Bell}}, \bibinfo {author} {\bibfnamefont {A.~A.}\ \bibnamefont {Rawlinson}},\ and\ \bibinfo {author} {\bibfnamefont {R.~F.}\ \bibnamefont {Sawyer}},\ }\bibfield  {title} {\bibinfo {title} {{Speedup through entanglement: Many body effects in neutrino processes}},\ }\href {https://doi.org/10.1016/j.physletb.2003.08.035} {\bibfield  {journal} {\bibinfo  {journal} {Phys. Lett. B}\ }\textbf {\bibinfo {volume} {573}},\ \bibinfo {pages} {86} (\bibinfo {year} {2003})},\ \Eprint {https://arxiv.org/abs/hep-ph/0304082} {arXiv:hep-ph/0304082} \BibitemShut {NoStop}%
\bibitem [{\citenamefont {Friedland}\ and\ \citenamefont {Lunardini}(2003)}]{Friedland:2003eh}%
  \BibitemOpen
  \bibfield  {author} {\bibinfo {author} {\bibfnamefont {A.}~\bibnamefont {Friedland}}\ and\ \bibinfo {author} {\bibfnamefont {C.}~\bibnamefont {Lunardini}},\ }\bibfield  {title} {\bibinfo {title} {{Do many particle neutrino interactions cause a novel coherent effect?}},\ }\href {https://doi.org/10.1088/1126-6708/2003/10/043} {\bibfield  {journal} {\bibinfo  {journal} {JHEP}\ }\textbf {\bibinfo {volume} {10}},\ \bibinfo {pages} {043}},\ \Eprint {https://arxiv.org/abs/hep-ph/0307140} {arXiv:hep-ph/0307140} \BibitemShut {NoStop}%
\bibitem [{\citenamefont {Duan}\ \emph {et~al.}(2010)\citenamefont {Duan}, \citenamefont {Fuller},\ and\ \citenamefont {Qian}}]{Duan:2010bg}%
  \BibitemOpen
  \bibfield  {author} {\bibinfo {author} {\bibfnamefont {H.}~\bibnamefont {Duan}}, \bibinfo {author} {\bibfnamefont {G.~M.}\ \bibnamefont {Fuller}},\ and\ \bibinfo {author} {\bibfnamefont {Y.-Z.}\ \bibnamefont {Qian}},\ }\bibfield  {title} {\bibinfo {title} {{Collective Neutrino Oscillations}},\ }\href {https://doi.org/10.1146/annurev.nucl.012809.104524} {\bibfield  {journal} {\bibinfo  {journal} {Ann. Rev. Nucl. Part. Sci.}\ }\textbf {\bibinfo {volume} {60}},\ \bibinfo {pages} {569} (\bibinfo {year} {2010})},\ \Eprint {https://arxiv.org/abs/1001.2799} {arXiv:1001.2799 [hep-ph]} \BibitemShut {NoStop}%
\bibitem [{\citenamefont {Chakraborty}\ \emph {et~al.}(2016)\citenamefont {Chakraborty}, \citenamefont {Hansen}, \citenamefont {Izaguirre},\ and\ \citenamefont {Raffelt}}]{Chakraborty:2016yeg}%
  \BibitemOpen
  \bibfield  {author} {\bibinfo {author} {\bibfnamefont {S.}~\bibnamefont {Chakraborty}}, \bibinfo {author} {\bibfnamefont {R.}~\bibnamefont {Hansen}}, \bibinfo {author} {\bibfnamefont {I.}~\bibnamefont {Izaguirre}},\ and\ \bibinfo {author} {\bibfnamefont {G.}~\bibnamefont {Raffelt}},\ }\bibfield  {title} {\bibinfo {title} {{Collective neutrino flavor conversion: Recent developments}},\ }\href {https://doi.org/10.1016/j.nuclphysb.2016.02.012} {\bibfield  {journal} {\bibinfo  {journal} {Nucl. Phys. B}\ }\textbf {\bibinfo {volume} {908}},\ \bibinfo {pages} {366} (\bibinfo {year} {2016})},\ \Eprint {https://arxiv.org/abs/1602.02766} {arXiv:1602.02766 [hep-ph]} \BibitemShut {NoStop}%
\bibitem [{\citenamefont {Tamborra}\ and\ \citenamefont {Shalgar}(2021)}]{Tamborra:2020cul}%
  \BibitemOpen
  \bibfield  {author} {\bibinfo {author} {\bibfnamefont {I.}~\bibnamefont {Tamborra}}\ and\ \bibinfo {author} {\bibfnamefont {S.}~\bibnamefont {Shalgar}},\ }\bibfield  {title} {\bibinfo {title} {{New Developments in Flavor Evolution of a Dense Neutrino Gas}},\ }\href {https://doi.org/10.1146/annurev-nucl-102920-050505} {\bibfield  {journal} {\bibinfo  {journal} {Ann. Rev. Nucl. Part. Sci.}\ }\textbf {\bibinfo {volume} {71}},\ \bibinfo {pages} {165} (\bibinfo {year} {2021})},\ \Eprint {https://arxiv.org/abs/2011.01948} {arXiv:2011.01948 [astro-ph.HE]} \BibitemShut {NoStop}%
\bibitem [{\citenamefont {Volpe}(2023)}]{Volpe:2023met}%
  \BibitemOpen
  \bibfield  {author} {\bibinfo {author} {\bibfnamefont {M.~C.}\ \bibnamefont {Volpe}},\ }\href@noop {} {\bibinfo {title} {{Neutrinos from dense environments : Flavor mechanisms, theoretical approaches, observations, and new directions}}} (\bibinfo {year} {2023}),\ \Eprint {https://arxiv.org/abs/2301.11814} {arXiv:2301.11814 [hep-ph]} \BibitemShut {NoStop}%
\bibitem [{\citenamefont {Cervia}\ \emph {et~al.}(2019)\citenamefont {Cervia}, \citenamefont {Patwardhan}, \citenamefont {Balantekin}, \citenamefont {Coppersmith},\ and\ \citenamefont {Johnson}}]{Cervia:2019res}%
  \BibitemOpen
  \bibfield  {author} {\bibinfo {author} {\bibfnamefont {M.~J.}\ \bibnamefont {Cervia}}, \bibinfo {author} {\bibfnamefont {A.~V.}\ \bibnamefont {Patwardhan}}, \bibinfo {author} {\bibfnamefont {A.~B.}\ \bibnamefont {Balantekin}}, \bibinfo {author} {\bibfnamefont {t.~S.~N.}\ \bibnamefont {Coppersmith}},\ and\ \bibinfo {author} {\bibfnamefont {C.~W.}\ \bibnamefont {Johnson}},\ }\bibfield  {title} {\bibinfo {title} {{Entanglement and collective flavor oscillations in a dense neutrino gas}},\ }\href {https://doi.org/10.1103/PhysRevD.100.083001} {\bibfield  {journal} {\bibinfo  {journal} {Phys. Rev. D}\ }\textbf {\bibinfo {volume} {100}},\ \bibinfo {pages} {083001} (\bibinfo {year} {2019})},\ \Eprint {https://arxiv.org/abs/1908.03511} {arXiv:1908.03511 [hep-ph]} \BibitemShut {NoStop}%
\bibitem [{\citenamefont {Patwardhan}\ \emph {et~al.}(2021)\citenamefont {Patwardhan}, \citenamefont {Cervia},\ and\ \citenamefont {Balantekin}}]{Patwardhan:2021rej}%
  \BibitemOpen
  \bibfield  {author} {\bibinfo {author} {\bibfnamefont {A.~V.}\ \bibnamefont {Patwardhan}}, \bibinfo {author} {\bibfnamefont {M.~J.}\ \bibnamefont {Cervia}},\ and\ \bibinfo {author} {\bibfnamefont {A.~B.}\ \bibnamefont {Balantekin}},\ }\bibfield  {title} {\bibinfo {title} {{Spectral splits and entanglement entropy in collective neutrino oscillations}},\ }\href {https://doi.org/10.1103/PhysRevD.104.123035} {\bibfield  {journal} {\bibinfo  {journal} {Phys. Rev. D}\ }\textbf {\bibinfo {volume} {104}},\ \bibinfo {pages} {123035} (\bibinfo {year} {2021})},\ \Eprint {https://arxiv.org/abs/2109.08995} {arXiv:2109.08995 [hep-ph]} \BibitemShut {NoStop}%
\bibitem [{\citenamefont {Illa}\ and\ \citenamefont {Savage}(2023)}]{Illa:2022zgu}%
  \BibitemOpen
  \bibfield  {author} {\bibinfo {author} {\bibfnamefont {M.}~\bibnamefont {Illa}}\ and\ \bibinfo {author} {\bibfnamefont {M.~J.}\ \bibnamefont {Savage}},\ }\bibfield  {title} {\bibinfo {title} {{Multi-Neutrino Entanglement and Correlations in Dense Neutrino Systems}},\ }\href {https://doi.org/10.1103/PhysRevLett.130.221003} {\bibfield  {journal} {\bibinfo  {journal} {Phys. Rev. Lett.}\ }\textbf {\bibinfo {volume} {130}},\ \bibinfo {pages} {221003} (\bibinfo {year} {2023})},\ \Eprint {https://arxiv.org/abs/2210.08656} {arXiv:2210.08656 [nucl-th]} \BibitemShut {NoStop}%
\bibitem [{\citenamefont {Martin}\ \emph {et~al.}(2023{\natexlab{a}})\citenamefont {Martin}, \citenamefont {Roggero}, \citenamefont {Duan},\ and\ \citenamefont {Carlson}}]{Martin:2023ljq}%
  \BibitemOpen
  \bibfield  {author} {\bibinfo {author} {\bibfnamefont {J.~D.}\ \bibnamefont {Martin}}, \bibinfo {author} {\bibfnamefont {A.}~\bibnamefont {Roggero}}, \bibinfo {author} {\bibfnamefont {H.}~\bibnamefont {Duan}},\ and\ \bibinfo {author} {\bibfnamefont {J.}~\bibnamefont {Carlson}},\ }\bibfield  {title} {\bibinfo {title} {{Many-body neutrino flavor entanglement in a simple dynamic model}},\ }\Eprint {https://arxiv.org/abs/2301.07049} {arXiv:2301.07049 [hep-ph]}  (\bibinfo {year} {2023}{\natexlab{a}})\BibitemShut {NoStop}%
\bibitem [{\citenamefont {Martin}\ \emph {et~al.}(2023{\natexlab{b}})\citenamefont {Martin}, \citenamefont {Neill}, \citenamefont {Roggero}, \citenamefont {Duan},\ and\ \citenamefont {Carlson}}]{Martin:2023gbo}%
  \BibitemOpen
  \bibfield  {author} {\bibinfo {author} {\bibfnamefont {J.~D.}\ \bibnamefont {Martin}}, \bibinfo {author} {\bibfnamefont {D.}~\bibnamefont {Neill}}, \bibinfo {author} {\bibfnamefont {A.}~\bibnamefont {Roggero}}, \bibinfo {author} {\bibfnamefont {H.}~\bibnamefont {Duan}},\ and\ \bibinfo {author} {\bibfnamefont {J.}~\bibnamefont {Carlson}},\ }\bibfield  {title} {\bibinfo {title} {{Equilibration of quantum many-body fast neutrino flavor oscillations}},\ }\Eprint {https://arxiv.org/abs/2307.16793} {arXiv:2307.16793 [hep-ph]}  (\bibinfo {year} {2023}{\natexlab{b}})\BibitemShut {NoStop}%
\bibitem [{\citenamefont {Patwardhan}\ \emph {et~al.}(2023)\citenamefont {Patwardhan}, \citenamefont {Cervia}, \citenamefont {Rrapaj}, \citenamefont {Siwach},\ and\ \citenamefont {Balantekin}}]{Patwardhan:2022mxg}%
  \BibitemOpen
  \bibfield  {author} {\bibinfo {author} {\bibfnamefont {A.~V.}\ \bibnamefont {Patwardhan}}, \bibinfo {author} {\bibfnamefont {M.~J.}\ \bibnamefont {Cervia}}, \bibinfo {author} {\bibfnamefont {E.}~\bibnamefont {Rrapaj}}, \bibinfo {author} {\bibfnamefont {P.}~\bibnamefont {Siwach}},\ and\ \bibinfo {author} {\bibfnamefont {A.~B.}\ \bibnamefont {Balantekin}},\ }\bibinfo {title} {{Many-Body Collective Neutrino Oscillations: Recent Developments}},\ in\ \href {https://doi.org/10.1007/978-981-15-8818-1_126-1} {\emph {\bibinfo {booktitle} {{Handbook of Nuclear Physics}}}},\ \bibinfo {editor} {edited by\ \bibinfo {editor} {\bibfnamefont {I.}~\bibnamefont {Tanihata}}, \bibinfo {editor} {\bibfnamefont {H.}~\bibnamefont {Toki}},\ and\ \bibinfo {editor} {\bibfnamefont {T.}~\bibnamefont {Kajino}}}\ (\bibinfo {year} {2023})\ pp.\ \bibinfo {pages} {1--16},\ \Eprint {https://arxiv.org/abs/2301.00342} {arXiv:2301.00342 [hep-ph]} \BibitemShut {NoStop}%
\bibitem [{\citenamefont {Shalgar}\ and\ \citenamefont {Tamborra}(2023)}]{Shalgar:2023ooi}%
  \BibitemOpen
  \bibfield  {author} {\bibinfo {author} {\bibfnamefont {S.}~\bibnamefont {Shalgar}}\ and\ \bibinfo {author} {\bibfnamefont {I.}~\bibnamefont {Tamborra}},\ }\bibfield  {title} {\bibinfo {title} {{Do we have enough evidence to invalidate the mean-field approximation adopted to model collective neutrino oscillations?}},\ }\href {https://doi.org/10.1103/PhysRevD.107.123004} {\bibfield  {journal} {\bibinfo  {journal} {Phys. Rev. D}\ }\textbf {\bibinfo {volume} {107}},\ \bibinfo {pages} {123004} (\bibinfo {year} {2023})},\ \Eprint {https://arxiv.org/abs/2304.13050} {arXiv:2304.13050 [astro-ph.HE]} \BibitemShut {NoStop}%
\bibitem [{\citenamefont {Johns}(2023)}]{Johns:2023ewj}%
  \BibitemOpen
  \bibfield  {author} {\bibinfo {author} {\bibfnamefont {L.}~\bibnamefont {Johns}},\ }\bibfield  {title} {\bibinfo {title} {{Neutrino many-body correlations}},\ }\Eprint {https://arxiv.org/abs/2305.04916} {arXiv:2305.04916 [hep-ph]}  (\bibinfo {year} {2023})\BibitemShut {NoStop}%
\bibitem [{\citenamefont {Kersten}\ and\ \citenamefont {Smirnov}(2016)}]{Kersten:2015kio}%
  \BibitemOpen
  \bibfield  {author} {\bibinfo {author} {\bibfnamefont {J.}~\bibnamefont {Kersten}}\ and\ \bibinfo {author} {\bibfnamefont {A.~Y.}\ \bibnamefont {Smirnov}},\ }\bibfield  {title} {\bibinfo {title} {{Decoherence and oscillations of supernova neutrinos}},\ }\href {https://doi.org/10.1140/epjc/s10052-016-4187-5} {\bibfield  {journal} {\bibinfo  {journal} {Eur. Phys. J. C}\ }\textbf {\bibinfo {volume} {76}},\ \bibinfo {pages} {339} (\bibinfo {year} {2016})},\ \Eprint {https://arxiv.org/abs/1512.09068} {arXiv:1512.09068 [hep-ph]} \BibitemShut {NoStop}%
\bibitem [{\citenamefont {Hannestad}\ \emph {et~al.}(2006)\citenamefont {Hannestad}, \citenamefont {Raffelt}, \citenamefont {Sigl},\ and\ \citenamefont {Wong}}]{Hannestad:2006nj}%
  \BibitemOpen
  \bibfield  {author} {\bibinfo {author} {\bibfnamefont {S.}~\bibnamefont {Hannestad}}, \bibinfo {author} {\bibfnamefont {G.~G.}\ \bibnamefont {Raffelt}}, \bibinfo {author} {\bibfnamefont {G.}~\bibnamefont {Sigl}},\ and\ \bibinfo {author} {\bibfnamefont {Y.~Y.~Y.}\ \bibnamefont {Wong}},\ }\bibfield  {title} {\bibinfo {title} {{Self-induced conversion in dense neutrino gases: Pendulum in flavour space}},\ }\href {https://doi.org/10.1103/PhysRevD.74.105010} {\bibfield  {journal} {\bibinfo  {journal} {Phys. Rev. D}\ }\textbf {\bibinfo {volume} {74}},\ \bibinfo {pages} {105010} (\bibinfo {year} {2006})},\ \bibinfo {note} {[Erratum: Phys.Rev.D 76, 029901 (2007)]},\ \Eprint {https://arxiv.org/abs/astro-ph/0608695} {arXiv:astro-ph/0608695} \BibitemShut {NoStop}%
\bibitem [{\citenamefont {Raffelt}\ and\ \citenamefont {Sigl}(2007)}]{Raffelt:2007yz}%
  \BibitemOpen
  \bibfield  {author} {\bibinfo {author} {\bibfnamefont {G.~G.}\ \bibnamefont {Raffelt}}\ and\ \bibinfo {author} {\bibfnamefont {G.}~\bibnamefont {Sigl}},\ }\bibfield  {title} {\bibinfo {title} {{Self-induced decoherence in dense neutrino gases}},\ }\href {https://doi.org/10.1103/PhysRevD.75.083002} {\bibfield  {journal} {\bibinfo  {journal} {Phys. Rev. D}\ }\textbf {\bibinfo {volume} {75}},\ \bibinfo {pages} {083002} (\bibinfo {year} {2007})},\ \Eprint {https://arxiv.org/abs/hep-ph/0701182} {arXiv:hep-ph/0701182} \BibitemShut {NoStop}%
\bibitem [{\citenamefont {Pastor}\ \emph {et~al.}(2002)\citenamefont {Pastor}, \citenamefont {Raffelt},\ and\ \citenamefont {Semikoz}}]{Pastor:2001iu}%
  \BibitemOpen
  \bibfield  {author} {\bibinfo {author} {\bibfnamefont {S.}~\bibnamefont {Pastor}}, \bibinfo {author} {\bibfnamefont {G.~G.}\ \bibnamefont {Raffelt}},\ and\ \bibinfo {author} {\bibfnamefont {D.~V.}\ \bibnamefont {Semikoz}},\ }\bibfield  {title} {\bibinfo {title} {{Physics of synchronized neutrino oscillations caused by selfinteractions}},\ }\href {https://doi.org/10.1103/PhysRevD.65.053011} {\bibfield  {journal} {\bibinfo  {journal} {Phys. Rev. D}\ }\textbf {\bibinfo {volume} {65}},\ \bibinfo {pages} {053011} (\bibinfo {year} {2002})},\ \Eprint {https://arxiv.org/abs/hep-ph/0109035} {arXiv:hep-ph/0109035} \BibitemShut {NoStop}%
\end{thebibliography}%
\end{document}